\documentclass[letterpaper,aip,jcp,numerical,reprint,10pt]{revtex4-1}
\usepackage[latin9]{inputenc}
\usepackage{color}
\usepackage{amsmath}
\usepackage{amssymb}
\usepackage{graphicx}
\usepackage{esint}

\makeatletter

\pdfpageheight\paperheight
\pdfpagewidth\paperwidth

\usepackage{footmisc}

\makeatother

\begin{document}

\title{On the Accuracy of the Padé--Resummed Master Equation Approach to
Dissipative Quantum Dynamics}

\author{Hsing-Ta Chen}

\affiliation{Department of Chemistry, Columbia University, 3000 Broadway, New
York, New York 10027, USA}

\author{Timothy C. Berkelbach}

\affiliation{Princeton Center for Theoretical Science, Princeton University, Princeton,
New Jersey 08544, USA}

\author{David R. Reichman}

\affiliation{Department of Chemistry, Columbia University, 3000 Broadway, New
York, New York 10027, USA}

\date{\today}
\begin{abstract}
Well--defined criteria are proposed for assessing the accuracy of
quantum master equations whose memory functions are approximated by
Padé resummation of the first two moments in the electronic coupling.
These criteria partition the parameter space into distinct levels
of expected accuracy, ranging from quantitatively accurate regimes
to regions of parameter space where the approach is not expected to
be applicable. Extensive comparison of Padé--resummed master equations
with numerically exact results in the context of the spin--boson model
demonstrate that the proposed criteria correctly demarcate the regions
of parameter space where the Padé approximation is reliable. The applicability
analysis we present is not confined to the specifics of the Hamiltonian
under consideration and should provide guidelines for other classes
of resummation techniques.
\end{abstract}
\maketitle

\section{Introduction\label{sec:Introduction}}

\textit{}

\textit{}The study of open quantum systems is among the most active
areas in condensed matter science.\cite{Leggett1987} Under the rubric
of dissipative quantum dynamics falls topics ranging from electron
and energy transfer\cite{Engel2007,Lee2007,Panitchayangkoon2010,Collini2009,Bredas2009}
to singlet fission dynamics in condensed media.\cite{Smith2010,Teichen2012a,Berkelbach2013a,Berkelbach2013,Berkelbach2014}
\textit{}The theoretical treatment of the dynamics of open quantum
systems is challenging due to the large number of degrees of freedom
and energy scales in such problems. Numerically exact approaches\cite{Makarov1994a,Makri1995,Makri1995a,Makri1995b,Mak2007,Wang2001,Wang2003,Tanimura1989,Ishizaki2009a,Strumpfer2012,Zhang2016}
are generally limited to idealized models, while approximate approaches\cite{Tully1998,Tully2012,Thoss2001,Berkelbach2012a,Montoya-Castillo2015}
are often limited by issues of both accuracy and scalablility. \textit{}Thus,
the search for methods that are reliable, general, and numerically
efficient continues at the forefront of theoretical chemistry and
physics.

\textit{}

Schemes based on projection operator techniques\cite{Grabert} and
generalized quantum master equations (GQMEs) have been used both to
design successful approximate approaches and as a platform to develop
numerically exact methods.\cite{Bloch1957,Tanimura1989,REDFIELD1965,weiss1999,Nitzan}\textit{
}The projection operator technique partitions the Hilbert space into
system and bath subspaces, leading to the derivation of GQME for the
system subspace which accounts for the bath's dynamical influence
on the system via a memory kernel. \textit{}Exact and approximate
techniques for the evaluation of the memory kernel have been developed
that make use of perturbation theories,\cite{Leggett1987,Wurger1997,Zhang1998,Yang2002,Ishizaki2009}
resummation techniques,\cite{Sparpaglione1988,Sparpaglione1988a,Golosov2001,Golosov2001a,Golosov2004,Mavros2014}
and self--consistent expansions.\cite{Shi2003,Shi2004,Cohen2011a,Cohen2013b}
\textit{}Recent progress afforded by these methods has illustrated
several advantages of the GQME scheme. \textit{}First, the memory
kernel may decay on a shorter timescale than the system dynamics under
study, so that approximate memory kernels may yield more accurate
dynamics than would be obtained by direct simulation of the system
dynamics using the same level of approximation. \textit{}Second,
the GQME scheme is general enough to treat realistic anharmonic baths\cite{Shi2004,Golosov2001a}
and arbitrary system--bath coupling.\cite{Shi2004} \textit{}Finally,
the flexibility of different projection operator formulations allows
for facile extension to more general situations, such as nonequilibrium
initial preparation,\cite{Evans1995,Coalson1994} as well as more
complex correlation functions.\cite{MontoyaCastillo2016} \textit{}However,
despite these notable results, it remains a difficult task to accurately
calculate memory kernels in many regimes of general quantum dissipative
systems.

\textit{}

\textit{}The Padé resummation approach approximates the memory kernel
as an infinite resummation based on the kernel's second and fourth
moments.\cite{Basdevant1972,Sparpaglione1988} At the expense of fourth--order
perturbation theory in the electronic coupling, the Padé--resummed
GQME is capable of producing an accuracy that exceeds that of simple
perturbation theory for the spin--boson model,\cite{Golosov2004}
and resummation of higher order kernels provide quantitative corrections.\cite{Gong2015}
\textit{}Recently, however, it has been demonstrated that this approach
can lead to unphysical, divergent dynamics in the strong electronic
coupling regime,\cite{Mavros2014} \textit{}and the applicability
and accuracy of the Padé approximations throughout the entire parameter
space is still difficult to evaluate.\textit{ }The aim of the present
work is to provide feasible estimates of the applicability based on
analysis of the Padé approximation itself.

\textit{}

We propose well--defined criteria in terms of the kernel's second
and fourth moments that correspond to conditions leading to ``physically
reasonable'' results within the Padé resummation scheme. To examine
the proposed criteria, we perform systematic benchmark comparisons
of Padé--resummation with numerically exact results for a prototypical
dissipative open quantum system, namely the spin--boson model with
a Debye spectral density.\textit{ }The proposed criteria divide the
parameter space into subspaces associated with different levels of
accuracy, and we confirm that the systematic comparison of population
dynamics with exact results clearly demarcate when the approach should
provide quantitatively reliable results. It should be noted that
the proposed criteria are not limited to the spin--boson model, but
are generally applicable for estimating the accuracy of Padé--resummed
memory kernels for generic open quantum systems. In addition, the
present work may provide guidelines for the applicability of other
types of resummation techniques, such as the Landau--Zener resummation.\cite{Mavros2014}

\textit{}

The outline of the paper is as follows. We present in Sec.~\ref{sec:Pade-resummed_GQME}
a brief review of the nonequilibrium Padé--resummed GQME approach
to a generic open quantum system. In Sec.~\ref{sec:Applicability_analysis_and_criteria},
we analyze the Padé resummation and define the criteria for the validity
of the approximation. We apply the proposed criteria to the spin--boson
model in Sec.~\ref{sec:IV. Results_for_SB_model} and show the correspondence
of the different regions of the applicability phase diagrams with
exactly computed population dynamics. In Sec.~\ref{sec:Conclusions},
we conclude.

\section{The Padé resummed GQME approach\label{sec:Pade-resummed_GQME}}

We consider an open quantum system whose Hamiltonian takes the form,
$\hat{H}=\hat{H}_{s}+\hat{H}_{b}+\hat{V}$, where $\hat{H}_{s}$ and
$\hat{H}_{b}$ correspond to the system and bath Hamiltonians, respectively,
and $\hat{V}$ denotes the system--bath coupling. We denote the quantum
states of the system by the kets $\left|j\right\rangle $ and the
bath density operator by $\hat{\rho}$. It is convenient to adopt
the Liouville space notation\cite{Sparpaglione1988,Sparpaglione1988a}
for the total density operator, $\hat{W}\equiv|W\rangle\rangle$,
and define the product $\langle\langle A|B\rangle\rangle\equiv\text{Tr}_{s}\text{Tr}_{b}\{A^{\dagger}B\}$
where $\text{Tr}_{s}$ and $\text{Tr}_{b}$ are partial traces over
the states of the system and bath, respectively. Time evolution of
the density operator is governed by the Liouville--von Neumann equation
\begin{equation}
\frac{d}{dt}|W(t)\rangle\rangle=-i\mathcal{L}|W(t)\rangle\rangle,\label{eq:Liouville}
\end{equation}
where the Liouville super--operator (the Liouvillian) is defined by
$\mathcal{L}|W(t)\rangle\rangle=[\hat{H},\hat{W}(t)]$ and we set
$\hbar=1$ throughout this paper. The reduced density matrix of the
system can be written as $\sigma_{jk}(t)=\text{Tr}_{b}\left\{ \left|k\right\rangle \left\langle j\right|W(t)\right\} =\langle\langle jk|W(t)\rangle\rangle$
where the Liouville state is given by $|jk\rangle\rangle=\left|j\right\rangle \left\langle k\right|\otimes\hat{1}$
and $\hat{1}$ is the unit operator for the bath. Then we can denote
the population dynamics as 
\begin{equation}
P_{j}(t)=\langle\langle j|W(t)\rangle\rangle,
\end{equation}
 where the diagonal elements are expressed as $|jj\rangle\rangle\rightarrow|j\rangle\rangle$
for simplicity.

We implement the standard projection operator technique\cite{Zwanzig1961}
via the super--operator 
\begin{equation}
\mathcal{P}=\sum_{j}|j\rho_{j}\rangle\rangle\langle\langle j|
\end{equation}
where $|j\rho_{j}\rangle\rangle=\left|j\right\rangle \left\langle j\right|\otimes\hat{\rho}_{j}$
and the bath density operator $\hat{\rho}_{j}$ is taken to be in
equilibrium in the electronic state $\left|j\right\rangle $. The
projected version of Eq.~\eqref{eq:Liouville} yields the GQME for
the population of the $j$-th state, 
\begin{equation}
\frac{d}{dt}P_{j}(t)=\mathcal{I}_{j}(t)-{\color{red}\sum_{k}}\int_{0}^{t}d\tau\mathcal{K}_{jk}(t-\tau)P_{k}(\tau),\label{eq: GQME_time}
\end{equation}
where the memory kernel matrix is 
\begin{equation}
\mathcal{K}_{jk}(t)=\langle\langle j|\mathcal{P}\mathcal{L}e^{-i\mathcal{Q}\mathcal{L}t}\mathcal{Q}\mathcal{L}|k\rho_{k}\rangle\rangle,\label{eq: Kernel_time}
\end{equation}
and the inhomogeneous terms are given by
\begin{equation}
\mathcal{I}_{j}(t)=-i\langle\langle j|\mathcal{P}\mathcal{L}e^{-i\mathcal{Q}\mathcal{L}t}\mathcal{Q}|W(0)\rangle\rangle,\label{eq: Inhomo_time}
\end{equation}
with $\mathcal{Q}=1-\mathcal{P}$. The inhomogeneous terms result
from the fact that the initial condition for the total density operator
will generally satisfy $\mathcal{Q}|W(0)\rangle\rangle\neq0$. For
cases $\mathcal{Q}|W(0)\rangle\rangle=0$, $\mathcal{I}_{j}(t)=0$.
In the frequency domain, Eq.~\eqref{eq: GQME_time} can be transformed
from an integro--differential equation into the algebraic form 
\begin{equation}
sp_{j}(s)=p_{j}(t=0)+I_{j}(s)-\sum_{k}K_{jk}(s)p_{k}(s)\label{eq:algebraic_master_equation}
\end{equation}
with the use of the one--side Laplace transformation, $f(s)=\int_{0}^{\infty}e^{-st}F(t)dt$,
where $s$ is a complex number. It should be noted that calculation
of the memory kernel matrix and the inhomogeneous terms is difficult
in part because dynamical evolution involves a projected propagator
$e^{-i\mathcal{Q}\mathcal{L}t}$.

To approximate the projected propagator, one can carry out a perturbation
treatment with respect to a perturbation $\hat{H}'$ and an unperturbed
Hamiltonian $\hat{H}_{0}=\hat{H}-\hat{H}'$. The Liouvillian can be
decomposed as $\mathcal{L}=\mathcal{L}_{0}+\mathcal{L}'$ and Eq.~\eqref{eq: Kernel_time}
and \eqref{eq: Inhomo_time} can be expanded in terms of $\mathcal{L}'$.
As a result, the memory matrix and the inhomogeneous terms in frequency
domain can be expressed as a moment expansion $K_{jk}(s)=\sum_{n=1}^{\infty}K_{jk}^{(2n)}(s)$
and $I_{j}(s)=\sum_{n=1}^{\infty}I_{j}^{(2n)}(s)$ with

\begin{equation}
K_{jk}^{(2n)}(s)=\langle\langle j|\left[\mathcal{L}'G_{0}(s)\mathcal{L}'G_{0}(s)\mathcal{Q}\right]^{n-1}\mathcal{L}'G_{0}(s)\mathcal{L}'|k\rho_{k}\rangle\rangle,\label{eq:K_2n_moment}
\end{equation}
and 
\begin{equation}
I_{j}^{(2n)}(s)=\langle\langle j|\mathcal{L}'G_{0}(s)[\mathcal{Q}\mathcal{L}'G_{0}(s)]^{2n-1}\mathcal{Q}|W(0)\rangle\rangle,
\end{equation}
where the unperturbed Green's function is $G_{0}(s)=\left(s+i\mathcal{L}_{0}\right)^{-1}$.
In practice, evaluating the $(2n)$-th order moment requires a Laplace
transformation for each time variable in a $(2n-1)$-time correlation
function. Clearly, the complexity of the terms in the moment expansion
grows quickly as the moment order increases.

The memory matrix and inhomogeneous terms may be approximated by
a Padé resummation using the second and fourth moments in the frequency
domain,
\begin{equation}
K_{jk}(s)\approx\frac{[K_{jk}^{(2)}(s)]^{2}}{K_{jk}^{(2)}(s)-K_{jk}^{(4)}(s)},\label{eq:memorey_pade}
\end{equation}
\begin{equation}
I_{j}(s)\approx\frac{[I_{j}^{(2)}(s)]^{2}}{I_{j}^{(2)}(s)-I_{j}^{(4)}(s)}.\label{eq:inhomo_pade}
\end{equation}
It should be noted that the Padé resummation is a rational expression
that include infinite orders of the perturbation $\hat{H}'$, but
the contributions of higher order than the fourth are approximated;
for example, $K_{jk}^{(6)}\approx\left[K_{jk}^{(4)}(s)\right]^{2}/K_{jk}^{(2)}(s)$.
The expressions of this section have been discussed before,\cite{Golosov2001a}
but a systematic analysis is lacking. We now focus precisely on this
issue.

\section{Applicability Analysis and criteria\label{sec:Applicability_analysis_and_criteria}}

The accuracy of the Padé resummation is unknown and depends on the
analyticity of an unknown function in the complex plane. Despite this
fundamental difficulty, we may estimate its validity via simple convergence
properties and physical requirements of the memory kernels. For simplicity
below, the criteria are expressed in terms of a single memory kernel
element $K(s)$, thereby suppressing the indices associated with memory
functions and inhomogeneous terms.

The Padé resummation can be viewed as a complex geometric series
which is expected to yield well--behaved results only within the disk
of convergence of the Laurent series that represents the expansion
in the complex plane. A necessary condition for such convergence is
$\|K^{(4)}(s)/K^{(2)}(s)\|<1$, for all $s\in\mathbb{C}$, where $K^{(n)}(s)$
is the $n$-th order expression given in Eq.~\eqref{eq:K_2n_moment}.
Since the inverse Laplace transformation is performed along the imaginary
axis $s=i\omega$, we restrict this condition to

\[
\text{(a)}\qquad\qquad\|K^{(4)}(i\omega)/K^{(2)}(i\omega)\|<1\text{~for real }\omega.\qquad\qquad
\]

The above condition is quite strict and may be relaxed by consideration
of the physical requirements of a generic memory kernel. Consider
the Laplace inversion via the contour integration of the Bromwich
integral $K(t)=\frac{1}{2\pi i}\int_{\mathcal{C}}K(s)e^{st}ds$, where
$\mathcal{C}$ is the vertical contour in the complex plane chosen
to include all singularities of $K(s)$ to the left of it. \cite{Arfken2013,Yonemoto2003}
The asymptotic physical behavior of the memory kernel dictates that
the poles of the Padé--resummed approximation \emph{cannot} have a
non--negative real part, otherwise the memory function would not be
guaranteed to decay to zero as $t\rightarrow\infty$. We assume that
the distribution of poles changes continuously and smoothly as the
parameters of the model changes, allowing us to focus on the imaginary
axis $s=i\omega$ and monitor the behavior of $K^{(4)}(i\omega)/K^{(2)}(i\omega)$.
In particular, the equality $\text{Re}[K^{(4)}(i\omega^{*})/K^{(2)}(i\omega^{*})]=1$
is a necessary (albeit not sufficient) condition for the occurrence
of a pole on the imaginary axis at $s=i\omega^{*}$, which obviate
the asymptotic decay of the memory kernel in real--time. We thus
propose a second condition
\[
\text{(b)}\qquad\qquad\text{Re}[K^{(4)}(i\omega)/K^{(2)}(i\omega)]<1\ \text{for real }\omega,\qquad\qquad
\]
which, excepting random occurrences, maintains that all poles are
confined to the left of the imaginary axis in the complex plane and
that the memory function is well behaved. Note that the first criterion
is stricter than the second since it corresponds to the interior of
a unit circle in the complex plane while the latter condition corresponds
to the entire complex plane to the left of the boundary at $\text{Re}[z]=1$.

These criteria are indeed crude because they rely on the the limited
information of the first two terms of an infinite expansion. We will
employ these conditions below as demarcation lines in parameter space
to gauge the reliability of the Padé approximation. As will be demonstrated,
the criteria provide robust if conservative guidelines for the domain
of applicability for Padé--resummed master equations.

\section{Results for the spin--boson model\label{sec:IV. Results_for_SB_model}}

\subsection{Padé--resummed GQME approach for the spin--boson model}

In this section, we examine the criteria suggested above via investigation
of the population dynamics in the spin--boson model. \textit{}The
spin--boson model is an idealization of an open quantum system which
contains most of the important generic features of more complicated
dissipative quantum systems while offering the advantage that numerically
exact algorithms exist for the calculation of its dynamics over a
wide range of parameter space.\cite{Tanimura1989,Makarov1994a,Makri1995,Thoss2001}
To produce benchmark results for the spin--boson model in this work,
we use the numerically exact hierarchical equations of motion (HEOM)
methodology in the Parallel Hierarchy Integrator (\texttt{PHI}).\cite{Strumpfer2012}

\textit{}We consider a two--level system with energy bias $\epsilon$
and constant electronic coupling $\Delta$

\begin{equation}
\hat{H}_{s}=\frac{\epsilon}{2}\hat{\sigma}_{z}+\Delta\hat{\sigma}_{x},
\end{equation}
and $\hat{\sigma}_{z}=\left|1\right\rangle \left\langle 1\right|-\left|2\right\rangle \left\langle 2\right|$
and $\hat{\sigma}_{x}=\left|1\right\rangle \left\langle 2\right|+\left|2\right\rangle \left\langle 1\right|$.
The two--level system is coupled to a bath consisting of an infinite
set of harmonic oscillators 
\begin{equation}
\hat{H}_{b}=\sum_{\alpha}\frac{\hat{P}_{\alpha}^{2}}{2}+\frac{1}{2}\omega_{\alpha}^{2}\hat{Q}_{\alpha}^{2}.
\end{equation}
Here, the frequency of the $\alpha$-th bath mode is $\omega_{\alpha}$,
while $\hat{P}_{\alpha}$, $\hat{Q}_{\alpha}$ refer to the mass--weighted
momenta and coordinates of the $\alpha$-th mode. The system--bath
coupling is taken to be of the form
\begin{equation}
\hat{V}=\hat{\sigma}_{z}\sum_{\alpha}c_{\alpha}\hat{Q}_{\alpha},
\end{equation}
where $c_{\alpha}$ is the coupling strength between the two--level
system and the $\alpha$-th harmonic oscillator. The spectral density
compactly describes the influence of the bath on the dynamics of the
system, and takes the form
\begin{equation}
J(\omega)=\frac{\pi}{2}\sum_{\alpha}\frac{c_{\alpha}^{2}}{\omega_{\alpha}}\delta(\omega-\omega_{\alpha}).
\end{equation}
In our study we choose the commonly used Debye spectral density\cite{Thoss2001}
\begin{equation}
J(\omega)=\frac{\lambda}{2}\frac{\omega\omega_{c}}{\omega^{2}+\omega_{c}^{2}},
\end{equation}
which is Ohmic at low frequency with a Lorentzian cutoff at high frequency.
The Debye spectral density is characterized by two parameters: the
characteristic bath frequency $\omega_{c}$, which represents the
average timescale of the bath response, and the reorganization energy
$\lambda=\sum_{\alpha}c_{\alpha}^{2}/2\omega_{\alpha}^{2}$, which
is a direct measure of the coupling strength between the system and
the bath. \textit{\emph{}}In electron--transfer theory, the Debye
spectral density is commonly used for the description of a solvent
environment with Debye dielectric relaxation (i.e. exponential in
time). 

Throughout this work, we employ an initial density operator for the
bath of the form 
\begin{equation}
\hat{\rho}_{0}=\frac{e^{-\beta\hat{H}_{b}}}{\text{Tr}_{b}\{e^{-\beta\hat{H}_{b}}\}},\label{eq:uncorrelated_bath}
\end{equation}
where $\beta=1/k_{B}T$ is the inverse temperature of the bath. This
initial condition corresponds to thermal equilibrium in the reservoir
in the absence of the system--bath coupling and is the initial density
operator of relevance for the description of an impulsive Franck--Condon
excitation. 

We implement a commonly used projection operator of the form,\cite{Sparpaglione1988,Sparpaglione1988a}

\begin{equation}
\mathcal{P}=|1\rho_{1}\rangle\rangle\langle\langle1|+|2\rho_{2}\rangle\rangle\langle\langle2|,\label{eq:projection_operator}
\end{equation}
where fully--dressed equilibrium bath density operators of the form
\begin{equation}
\hat{\rho}_{j}=\frac{e^{-\beta\hat{H}_{j}}}{\text{Tr}_{b}\{e^{-\beta\hat{H}_{j}}\}}
\end{equation}
are employed with $\hat{H}_{j}=\pm(\frac{\epsilon}{2}+\sum_{\alpha}c_{\alpha}\hat{Q}_{\alpha})+\hat{H}_{b}$
($+$ for $1$ and $-$ for $2$). Note that with the use of the projector~\eqref{eq:projection_operator},
factorized initial conditions with an uncorrelated bath~\eqref{eq:uncorrelated_bath}
will necessitate the evolution of inhomogeneous terms~\eqref{eq:inhomo_pade}
in the GQME. The second--order moments of the memory kernels (\textbf{$K_{jk}^{(2)}$})
result in an expression equivalent to the noninteracting blip approximation
(NIBA).\cite{weiss1999} We carry out the time integrations of the
memory kernels and the inhomogeneous terms by the techniques outlined
in Ref.~\onlinecite{Golosov2004} and the same Gaussian quadrature
subroutine (\texttt{DCUTRI}).\cite{Berntsen1991} The population dynamics
of the Padé--resummed GQME is calculated via the accuracy--improved
numerical method for Laplace inversion.\cite{Yonemoto2003} 

For this spin--boson model, the Padé--resummed GQME approach has
lead to population dynamics in near perfect agreement with numerically
exact simulations.\cite{Golosov2001,Golosov2004} On the other hand,
Van Voorhis and coworkers have shown the breakdown of the Padé--resummed
GQME approach in the strong electronic coupling region.\cite{Mavros2014}
Our goal in the following is to systematically delineate the regime
of validity of the approach based on the criteria of Sec~\ref{sec:Applicability_analysis_and_criteria}.

\begin{figure}[h]
\includegraphics{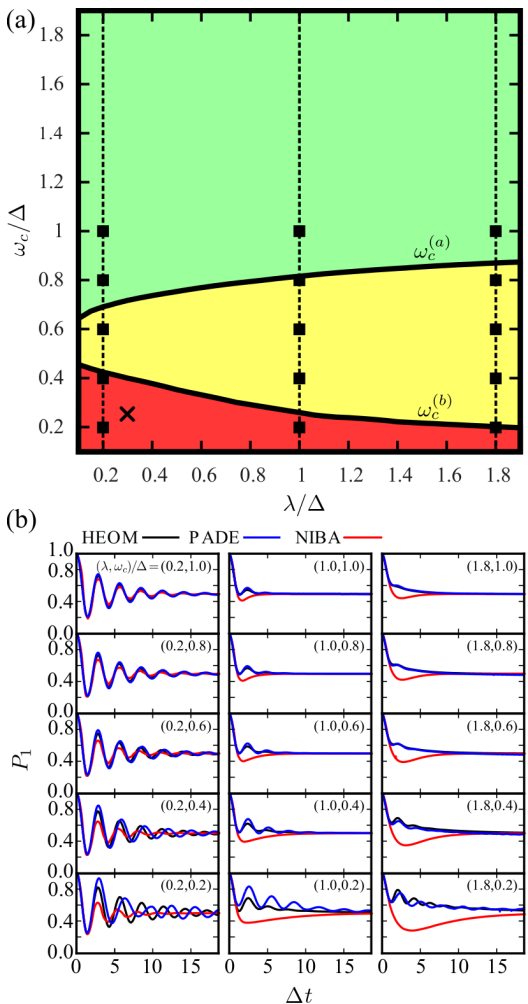} \caption{Parameter space diagram for the spin--boson model with zero bias ($\epsilon=0$)
and at high temperature ($k_{B}T=2\Delta$). The critical frequencies
$\omega_{c}^{(a)}$ and $\omega_{c}^{(b)}$ are indicated as functions
of $\lambda$. The green region ($\omega_{c}>\omega_{c}^{(a)}$) is
the regime where dynamics are expected to be quantitatively accurate,
the yellow region ($\omega_{c}^{(b)}<\omega_{c}<\omega_{c}^{(a)}$)
is the regime where dynamics are expected to be semi--quantitatively
accurate and the red region ($\omega_{c}<\omega_{c}^{(b)}$) is the
regime where the Padé--resummed approach is expected to be unreliable
or even unstable. The lower panels are the corresponding population
dynamics along the vertical cuts (indicated as solid squares connected
by dashed lines) calculated by the HEOM approach (red solid lines),
Padé--resummed GQME (PADE, green dash lines), and NIBA (blue doted
lines). The upper right label in each population dynamics panel denotes
the value of $(\lambda,\omega_{c})/\Delta$. The symbol $\boldsymbol{\times}$
in the phase diagram refers to the parameters corresponding to Fig.~1(d)
of Ref.~\onlinecite{Mavros2014}.\label{fig:unbiased_high_temperature} }
\end{figure}

\begin{figure*}[t]
\begin{centering}
\includegraphics[scale=0.94]{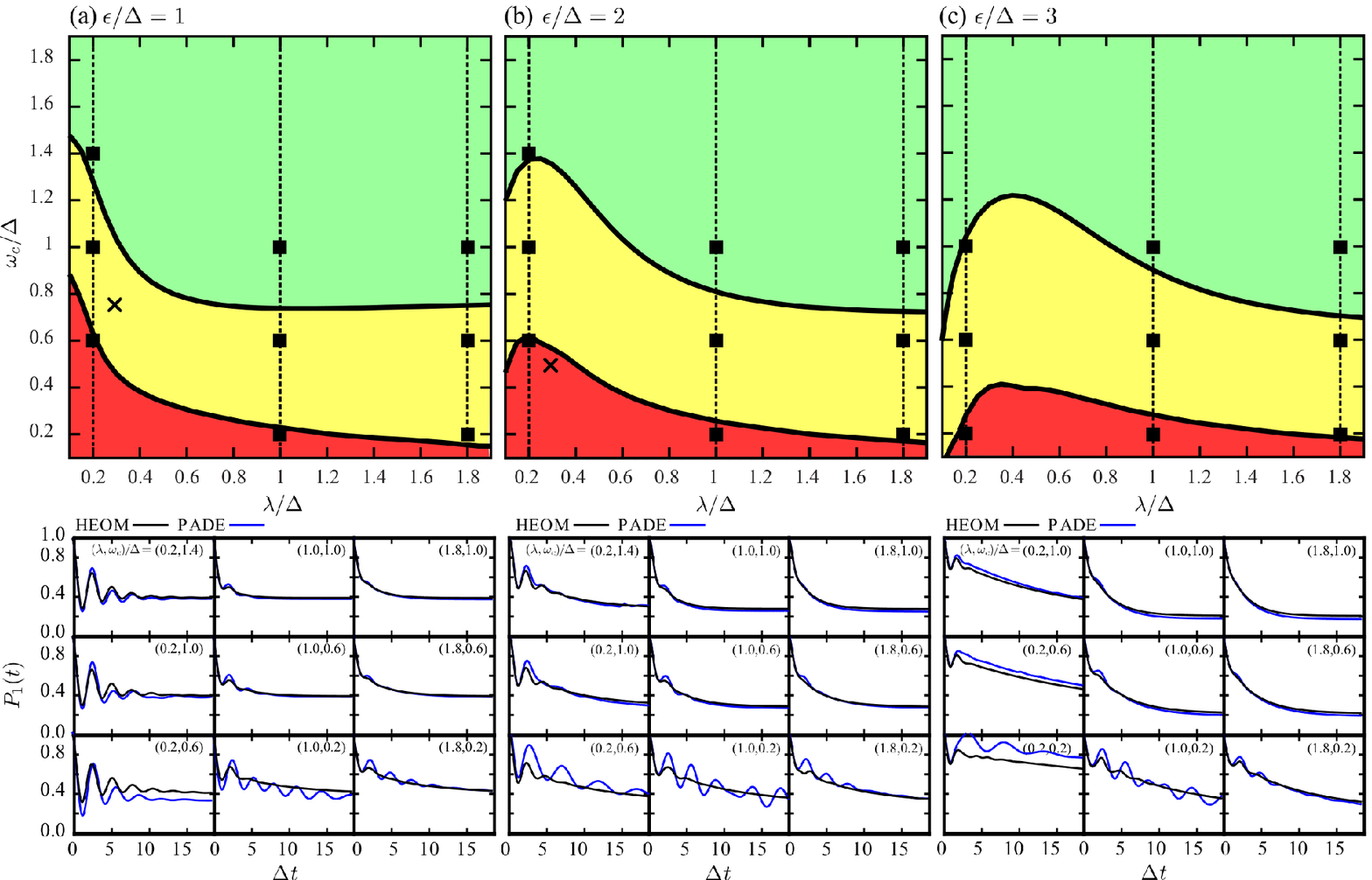}
\par\end{centering}

\caption{Parameter space diagram with increasing bias energies $\epsilon/\Delta=1,\ 2,\ 3$
at high temperatures ($k_{B}T=2\Delta$) for the spin--boson model.
The critical frequencies $\omega_{c}^{(a)}$ and $\omega_{c}^{(b)}$
are indicated as functions of $\lambda$ with color regions as in
Fig.~\ref{fig:unbiased_high_temperature}. The lower panels are the
corresponding population dynamics along the vertical cuts calculated
by the HEOM approach (red solid line) and the Padé--resummed GQME
(PADE, green dash line). The upper right label in each population
dynamics plot denotes the value of $(\lambda,\omega_{c})/\Delta$.
The symbol $\boldsymbol{\times}$ in panel (a) refers to the parameters
corresponding to Fig.~3(b) of Ref.~\onlinecite{Mavros2014}, while
that in panel (b) corresponds to Fig.~4(b) of Ref.~\onlinecite{Mavros2014}.\label{fig:bias_energy_dependence} }
\end{figure*}

\subsection{Parameter Space Diagrams}

The model we study in this work can be parametrized by five independent
energy scales. We use the electronic coupling $\Delta$ as the unit
of energy so that four dimensionless parameters characterize the parameter
space. These are: the electronic bias $\epsilon/\Delta$, the reorganization
energy $\lambda/\Delta$, the bath's characteristic frequency $\omega_{c}/\Delta$,
and the thermal energy of the bath $k_{B}T/\Delta$.

To systematically scan parameter space, we consider variation in the
scaled $\omega_{c}$--$\lambda$ plane for different scaled temperature
and bias cuts. It is expected that, for a given system--bath coupling
$\lambda$, smaller values of $\omega_{c}/\Delta$ render the Padé
approximation less accurate due to the fact that the perturbation
series is ordered by $\Delta$. Therefore, we define critical characteristic
frequencies, $\omega_{c}^{(a)}(\lambda)$ and $\omega_{c}^{(b)}(\lambda)$,
as the \emph{lower} bound of scaled $\omega_{c}$ to satisfy the criteria
(a) and (b) for all elements of the memory kernels respectively. The
boundaries $\omega_{c}^{(a)}(\lambda)$ and $\omega_{c}^{(b)}(\lambda)$
are determined by the conditions that there exists a single imaginary
number $i\omega^{*}$ for which either
\[
\qquad\text{(a)}\qquad\quad\|K^{(4)}(i\omega^{*})/K^{(2)}(i\omega^{*})\|=1,\qquad\qquad\qquad
\]
or
\[
\qquad\text{(b)}\qquad\text{Re}[K^{(4)}(i\omega^{*})/K^{(2)}(i\omega^{*})]=1,\qquad\qquad\qquad
\]
is satisfied. The critical characteristic frequencies indicate the
boundaries of the proposed criteria that partition parameter space
into three distinct regions of different levels of accuracy.

Because $K(t)$ and $I(t)$ have similar structure that should decay
to zero after a transient time and the Padé approximation takes the
same form for both $K(s)$ and $I(s)$, we expect the proposed criteria
also apply to the inhomogeneous term. In fact, Refs.~\onlinecite{Golosov2001}--\onlinecite{Golosov2004}
have shown that the initial preparation of the bath, captured by the
inhomogeneous term, is crucial for obtaining the correct dynamics.
We only focus here on the memory kernel and expect the inhomogeneous
term have similar analytical behaviors.

\begin{figure*}
\includegraphics[scale=0.94]{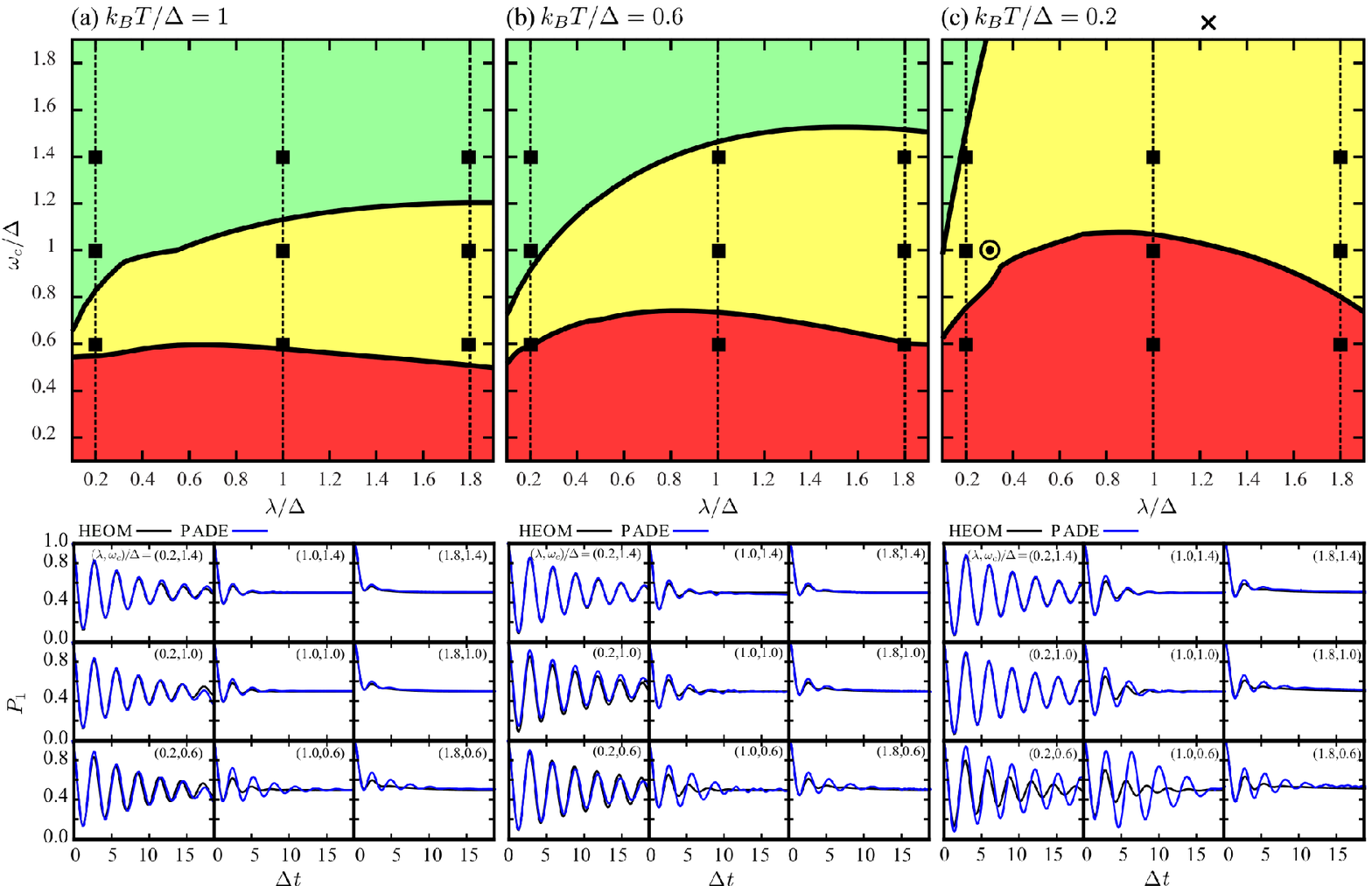}

\caption{Parameter space diagrams with zero bias energy ($\epsilon=0$) as
a function of decreasing temperature $k_{B}T/\Delta=1.0,\ 0.6,\ 0.2$
from left to right. The critical frequencies $\omega_{c}^{(a)}$ and
$\omega_{c}^{(b)}$ are indicated as functions of $\lambda$ with
color regions as in Fig.~\ref{fig:unbiased_high_temperature}. The
lower panels are the corresponding population dynamics along the vertical
cuts calculated by the HEOM approach (red solid line) and Padé--resummed
GQME (PADE, green dash line). The upper right label in each population
dynamics panel denotes the value of $(\lambda,\omega_{c})/\Delta$.\textcolor{blue}{{}
}The symbol $\boldsymbol{\times}$ in panel (c) refers to the parameters
corresponding to Fig.~2(b) of Ref.~\onlinecite{Mavros2014}. The
symbol $\odot$ indicates the same $(\lambda,\omega_{c})$ of Fig.~3(c)
of Ref.~\onlinecite{Mavros2014}, but with $\epsilon=0$.\label{fig:temperature_dependence} }
\end{figure*}

\textit{}For illustrative purposes, we show a phase diagram for an
unbiased ($\epsilon=0$), high temperature ($k_{B}T=2\Delta$) system
in Fig.~\ref{fig:unbiased_high_temperature} and the corresponding
population dynamics of selected points in parameter space calculated
by the HEOM, Padé and NIBA approaches in the lower panels. For this
example the three regions may be partitioned as:
\begin{enumerate}
\item $\omega_{c}>\omega_{c}^{(a)}$ (quantitatively accurate):

In this regime, the results of the Padé approach achieve almost perfect
agreement with the numerically exact results. This regime covers the
weak electronic coupling (non--adiabatic) limit ($\omega_{c}/\Delta\gg1$),
where the $\Delta$--perturbation based methods works well. We also
find that the Padé--resummed approach provides a significant improvement
over NIBA in the large system--bath coupling regime, as can be seen
in the upper panels of Fig.~\ref{fig:unbiased_high_temperature}
(b).

\item $\omega_{c}^{(b)}<\omega_{c}<\omega_{c}^{(a)}$ (semi--quantitatively
accurate):

The population dynamics of the Padé approach in this region are not
quite as accurate as in the ``quantitatively accurate'' regime,
but the Padé--resummed method still captures most of the important
features in a semi--quantitative manner, such as the long--lived oscillations
and dissipative relaxation. Since the electronic coupling is considered
to be intermediate, the NIBA results become worse while the Padé results
remain accurate.

\item $\omega_{c}<\omega_{c}^{(b)}$ (unreliable):

The discrepancies in the population dynamics between the Padé approach
and the HEOM generally become larger in this regime since the large
electronic coupling ($\Delta/\omega_{c}\gg1$) renders the perturbation
theory in $\Delta$ questionable. In this regime, the Padé approach
may lead to a shift of the oscillation frequency of the population
(see panels labeled by $(\lambda,\omega_{c})/\Delta=(0.2,0.4),\ (0.2,0.2),\ (1.8,0.2)$),
as well as overly coherent behavior (see the panel labeled by $(\lambda,\omega_{c})/\Delta=(1.0,0.2)$).
Extreme cases in the strong electronic coupling (adiabatic) limit
may cause the Padé resummation breakdown and result in unphysical
population dynamics. Importantly, the parameters of Fig.~1(d) of
Ref.~\onlinecite{Mavros2014} lie in the ``unreliable'' region
(labeled by $\boldsymbol{\times}$ in the phase diagram). In this
case the Padé--resummed approach yields unphysical population dynamics
for the long time behavior.\footnote{Despite qualitatively similar behaviors, we notice that our results
near the parameters marked by $\boldsymbol{\times}$ appear to be
more accurate than those of Ref.~\onlinecite{Mavros2014}. One possible
reason may be attributed to numerical errors of the FFT--based Laplace
inversion method of Honig and Hirdes.\cite{Honig1984} Here, we employ
a simple improved method proposed by Yonemoto et al.\cite{Yonemoto2003}
In addition, we note that Ref.~\onlinecite{Mavros2014} assumes $I_{j}(s)=0$
which may yield different population dynamics for short times.\label{fn:Despite-qualitatively-similar}}

\end{enumerate}
\textit{}

\subsection{Energetic bias dependence }

\textit{}The bias dependence of the parameter space phase diagram
is shown in Fig.~\ref{fig:bias_energy_dependence}, as well as the
corresponding population dynamics. We find that, as the energetic
bias grows, both critical frequencies increase in the low $\lambda$
region. Furthermore, in the region when $\omega_{c}<\omega_{c}^{(b)}$,
the Padé approach may lead to incorrect steady state population values
(see the panels labeled $(\lambda,\omega_{c})/\Delta=(0.2,0.6)$ for
$\epsilon=\Delta$ and $(\lambda,\omega_{c})/\Delta=(0.2,0.2)$ for
$\epsilon=3\Delta$) as well as an unphysical ``recoherence'' behavior
(namely the envelope of the population does not decay monotonically)
as illustrated in the panels $(\lambda,\omega_{c})/\Delta=(1.0,0.2)$
for all biases. This effect can be attributed to near singularities
in the approximate kernels when the Padé resummation does not satisfy
the criterion (b). The population dynamics in Fig.~3(b) and Fig.~4(b)
of Ref.~\onlinecite{Mavros2014} show qualitatively similar discrepancies
from exact calculations as illustrated here.\cite{Note1} The parameters
for these two cases (labeled as $(\times)$ in Fig.~\ref{fig:bias_energy_dependence})
lie in the expected regions of parameter space. 

\textit{}We find that $\omega_{c}^{(a)}$ and $\omega_{c}^{(b)}$
become insensitive to the energetic bias in the limit $\lambda\gg\epsilon$.
Since the reorganization processes dominate the incoherent decay in
this limit, the fluctuations induced by the energetic bias becomes
less important here. Hence, the boundaries of accuracy of the Padé--resummed
GQME approach do not change when system--bath coupling becomes very
large.

\subsection{Temperature dependence }

\textit{}

In general, the Padé--resummed GQME approach becomes less accurate
for lower temperature baths. Fig.~\ref{fig:temperature_dependence}
shows that, as the temperature decreases, the critical frequencies
increase significantly throughout the entire range of reorganization
energies. This may be explained by the fact that the bath degrees
of freedom progressively populate lower frequency modes as temperature
decreases, rendering $\Delta$ relatively larger with respect to the
participating low--frequency modes. However, the Padé approach can
still properly capture the dynamical effect of the bath and yield
qualitatively reasonable results in the semi--quantitatively accurate
region. 

In the regions of lower accuracy, the Padé approach tends to overestimate
the coherent oscillations. In addition, the coherent oscillations
are generally shifted toward lower frequencies. In addition, we observe
spurious recoherence in the panel labeled $(\lambda,\omega_{c})/\Delta=(1.0,0.6)$
for $k_{B}T=0.2\Delta$. Once again the most sever deviations from
exact calculation are found in the region $\omega_{c}<\omega_{c}^{(b)}(\lambda)$
as expected. 

The value $(\lambda,\omega_{c})/\Delta=(1.0,0.3)$ of Fig.~3(c) of
Ref.~\onlinecite{Mavros2014} is labeled ($\odot$) in panel (c)
of Fig.~\ref{fig:temperature_dependence}. However, note that the
values of the energetic bias are different in this comparison. As
discussed above, we expect both critical frequencies to increase in
the low $\lambda$ region as the value of bias grows. Hence, we infer
by this trend that the value of $(\lambda,\omega_{c})/\Delta$ in
the biased case should lie in the region of parameter space where
the Padé approach is expected to be unreliable.

\section{Conclusions\label{sec:Conclusions}}

\textit{}In this work we provide criteria to estimate the accuracy
and applicability of the nonequilibrium Padé--resummed GQME approach
to dissipative quantum dynamics. For the spin--boson model, the criteria
yield critical frequencies $\omega_{c}^{(a)}(\lambda)$ and $\omega_{c}^{(b)}(\lambda)$
that partition the parameter space into three distinct regions of
expected accuracy. One particularly significant outcome of our analysis
is the fact that the difficult intermediate coupling regime, where
all energy scales are comparable, falls frequently into a region of
parameter space where the Padé approach is expected to be accurate,
and indeed we find that the Padé--resummed GQME can still capture
significant features of population dynamics within this regime.\footnote{The criteria should be generally applicable in the larger reorganization
energy regime than we present here. In fact, the NIBA approach is
capable of producing quantitatively accurate results in the ``golden
rule'' regime where the reorganization energy is sufficiently larger
than the diabatic coupling ($\lambda\gg\Delta$). In addition, Fig~1~(a)~and~(b)
of Ref.~\onlinecite{Mavros2014} show that Padé GQME approach does
capture the dynamics well for large $\lambda/\Delta$. Therefore,
we expect the asymptotic behavior of the Padé GQME approach to be
as good as or better than the NIBA approach in this regime} When $\omega_{c}<\omega_{c}^{(b)}(\lambda)$, the Padé--resummed
GQME is demonstrated to often exhibit spurious long--time behavior,
overestimate oscillations with shifted frequencies, and display unphysical
recoherence. Overall, we find that the accuracy of the Padé resummation
is relatively insensitive to the system bias and reorganization energy,
but becomes worse with decreasing bath frequency and decreasing temperature.

The criteria of accuracy we propose is crude for several reasons.
First, it is only based on the analytic properties of the first two
moments of an infinite expansion. Second, even with regard to these
moments, we merely search for the boundaries in the complex plane
where a \emph{single} pole may obviate physical properties required
of generic memory functions. In this sense, the boundaries of accuracy
are conservative and we expect to see cases where the Padé approach
may still yield accurate results even if $\omega_{c}^{(b)}(\lambda)<\omega_{c}<\omega_{c}^{(a)}(\lambda)$
and even occasionally when $\omega_{c}<\omega_{c}^{(b)}(\lambda)$
. Indeed, we do find cases where exact calculations demonstrate that
the approximate results may be more accurate than expected. However,
overall we find that the trends predicted by the criteria of Sec.~\ref{sec:Applicability_analysis_and_criteria}
faithfully delineate the trends of accuracy of the Padé--resummed
generalized master equation approach.

\textit{}The proposed criteria should be valid for Padé resummations
used to approximate the memory kernels produced by other types of
projection operators, and our applicability analysis may provide guidelines
for assessing the domain of validity of other resummation techniques.
In particular, one can construct applicability phase diagrams for
other theories, such as the Landau--Zener resummation, leading to
an increased understanding of the domain of validity of complimentary
approaches. This line of investigation will be taken up in future
work.
\begin{acknowledgments}
We would like to thank Troy Van Voorhis, Michael Mavros, and Andrés
Montoya-Castillo for extensive discussions. This work was supported
by grant NSF CHE--1464802.
\end{acknowledgments}

\bibliographystyle{aipnum4-1}
\bibliography{Pade}

\end{document}